\documentclass[conference]{IEEEtran}
\IEEEoverridecommandlockouts

\usepackage{cite}
\usepackage[driverfallback=dvipdfm]{hyperref}
\usepackage{amsmath,amssymb,amsfonts}

\usepackage{algorithmic}
\usepackage[pdftex]{graphicx}
\usepackage{color}
\usepackage{textcomp}
\usepackage{xcolor}
\usepackage{url}
\usepackage{CJKutf8}
\usepackage{listings}
\makeatletter
\def\lst@CJK@XXX#1#2#3{%
  \lst@ifletter\else\lst@OutputOther\fi
  \lst@whitespacefalse
  \advance\lst@length\@ne
  \lst@AppendOther{\let\lst@hss\relax\csname CJK@\number`#1\endcsname{`#2}{`#3}}%
}%
\lst@AddToHook{SelectCharTable}{%
  \let\CJK@XXX\lst@CJK@XXX
}
\makeatother
\def\BibTeX{{\rm B\kern-.05em{\sc i\kern-.025em b}\kern-.08em
    T\kern-.1667em\lower.7ex\hbox{E}\kern-.125emX}}
\begin{document}

\title{A Learning Support Method for Multi-threaded Programs Using Trace Tables\\
}

\author{
\IEEEauthorblockN{Takumi Murata}
\IEEEauthorblockA{\textit{Electronics, Information and Media Engineering Major} \\
\textit{Graduate School of Engineering} \\
\textit{Nippon Institute of Technology}\\
Saitama, Japan \\
2228017@stu.nit.ac.jp}
\and
\IEEEauthorblockN{Hiroaki Hashiura \href{https://orcid.org/0000-0002-6325-4177}{\includegraphics[scale=0.08]{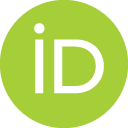}}}
\IEEEauthorblockA{\textit{Dept. of Data Science} \\
\textit{Faculty of Advanced Engineering} \\
\textit{Nippon Institute of Technology}\\
Saitama, Japan \\
hashiura@nit.ac.jp}
}

\maketitle

\begin{abstract}
Multi-threaded programs are expected to improve responsiveness and conserve resources by dividing an application process into multiple threads for concurrent processing.
However, due to scheduling and the interaction of multiple threads, their runtime behavior is more complex than that of single-threaded programs, making which makes debugging difficult unless the concepts specific to multi-threaded programs and the execution order of instructions can be understood.
In this paper, we propose a learning tool for multi-threaded programs using trace tables.
\end{abstract}


\section{Introduction}
In software development, debugging is an essential task for removing software faults.
It has been reported that debugging can consume approximately 30\% to 90\% of the total development time\cite{8109281}.
To address this issue, various methods have been proposed to improve debugging efficiency\cite{vedebug}\cite{BPGen}.

In programming education at universities, emphasis is placed on learning the concepts of algorithms and the syntax of programming languages.
However, programming novices have major problems when dealing with errors.
This poses a significant obstacle learning to code\cite{michaeli_current_2019}.
This indicates the importance of debugging in education and learning.
On the other hand, debugging has generally not been a heavily emphasized area in programming curricula.

One form of programming used in web applications and other software is multi-threaded programming.
This involves dividing a single application’s process into multiple threads to enable parallel processing, which can improve responsiveness and save resources by executing multiple processes simultaneously. However, due to scheduling and the interaction between multiple threads, the runtime behavior is often more complex than that of single-threaded programs.

multi-threaded programs perform processing while switching between threads.
Depending on the timing of these switches, bugs specific to multithreading, such as race conditions and deadlocks, can occur.
These multithreading-related bugs can sometimes have low reproducibility\cite{Oracle}, which makes debugging difficult with debuggers that cannot investigate the program’s past state.
Furthermore, it is necessary to understand concepts specific to multithreading and the execution order of threads.

Teaching multi-threaded programming to beginners is a difficult task, both for the teacher who has to find the best way to illustrate the concepts related to concurrency and for students who have to figure out what happens in their programs.
It has also been noted that when debugging multi-threaded programs, most students tend to rely on analysis and debugging techniques that have been successful in single-threaded programming, despite being warned against doing so\cite{malnati_jthreadspy_2007}.

In this paper, we describes the results of developing an educational tool for debugging multi-threaded programs.

\section{Method}
The debugger, which is widely used in program debugging, pauses the program at breakpoints to examine the state of variables and other elements.
Therefore, developers must, of course,  set breakpoints before executing the program.
To set these breakpoints, developers need to know which variables are causing the bugs.
Moreover, because programs that use a debugger usually cannot go backward; thus, if a developer wants to investigate the state before the breakpoint, they need to reset the breakpoint and rerun the debugging process.

A program debugging method is to use traces. 
Traces involves recording the events that occur during program execution (such as method execution details and variable assignments) sequentially and comprehensively. 
In debugging that uses traces, developers can identify the location of faults by examining the trace backward from the failure point, reading changes in the variables and control flows. 
This work is performed after the program execution has finished and has the advantage, compared to using a debugger, of not requiring predetermined stopping points for the program\cite{murata_trace}.

Therefore, debugging using traces is more effective than using breakpoints for debugging multi-threaded programs.

\subsection{Trace Table}
\begin{figure}[h]
    \begin{center}
        \includegraphics[width=7cm]{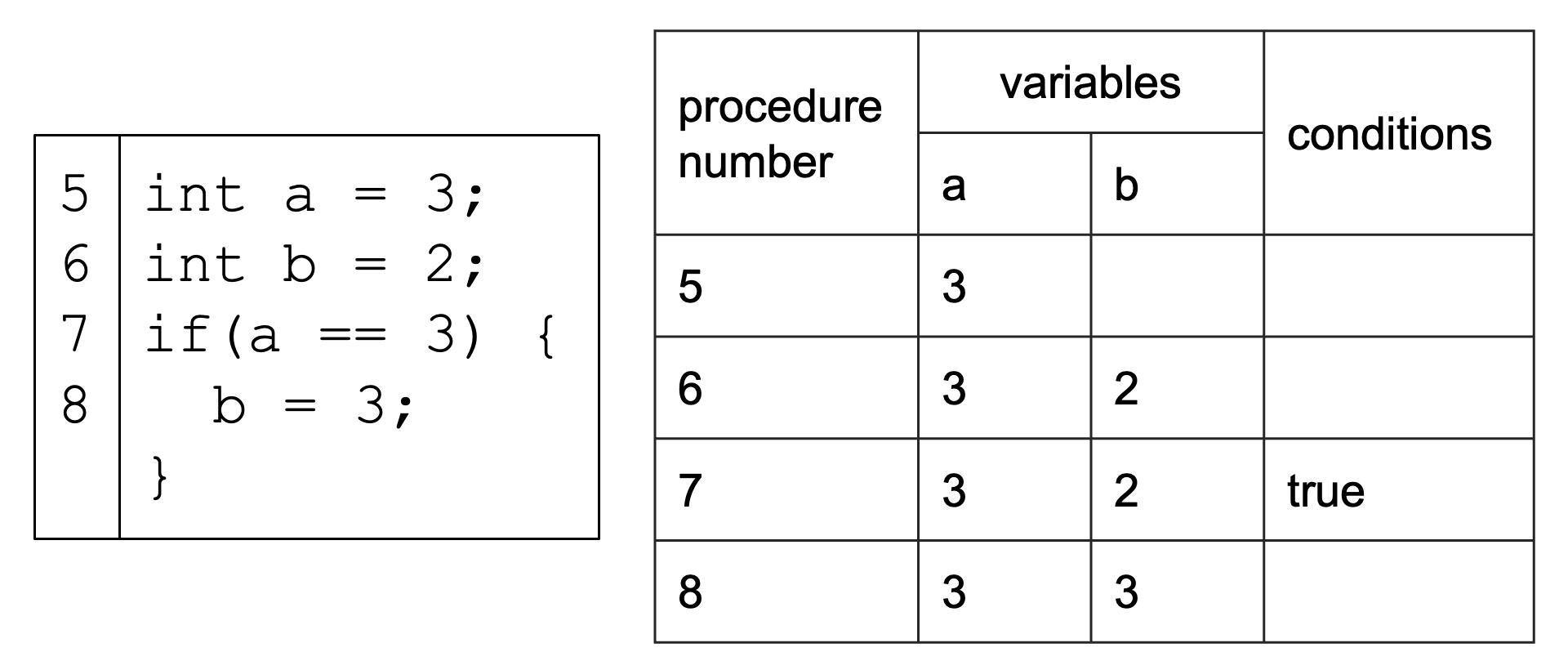}
        \caption{Example of trace table}\label{トレース表の例}
    \end{center}
\end{figure}

To perform debugging, it is necessary to accurately understand how the values of variables change in each line of the program.
One method to support learners in debugging during programming is to use a trace table.
A trace table\cite{ozaki_trace_table}, as shown in figure \ref{トレース表の例}, sequentially records the values of variables and other information at each step of program execution, including the procedure number, variable names, and variable values.
When using a trace table, if learners do not accurately understand the program's control flow, the content of each line, or the changes in variable values, these errors will be reflected in the trace table they create.

\subsection{Enhanced Trace Tables}
\begin{figure}[h]
    \begin{center}
        \includegraphics[width=7cm]{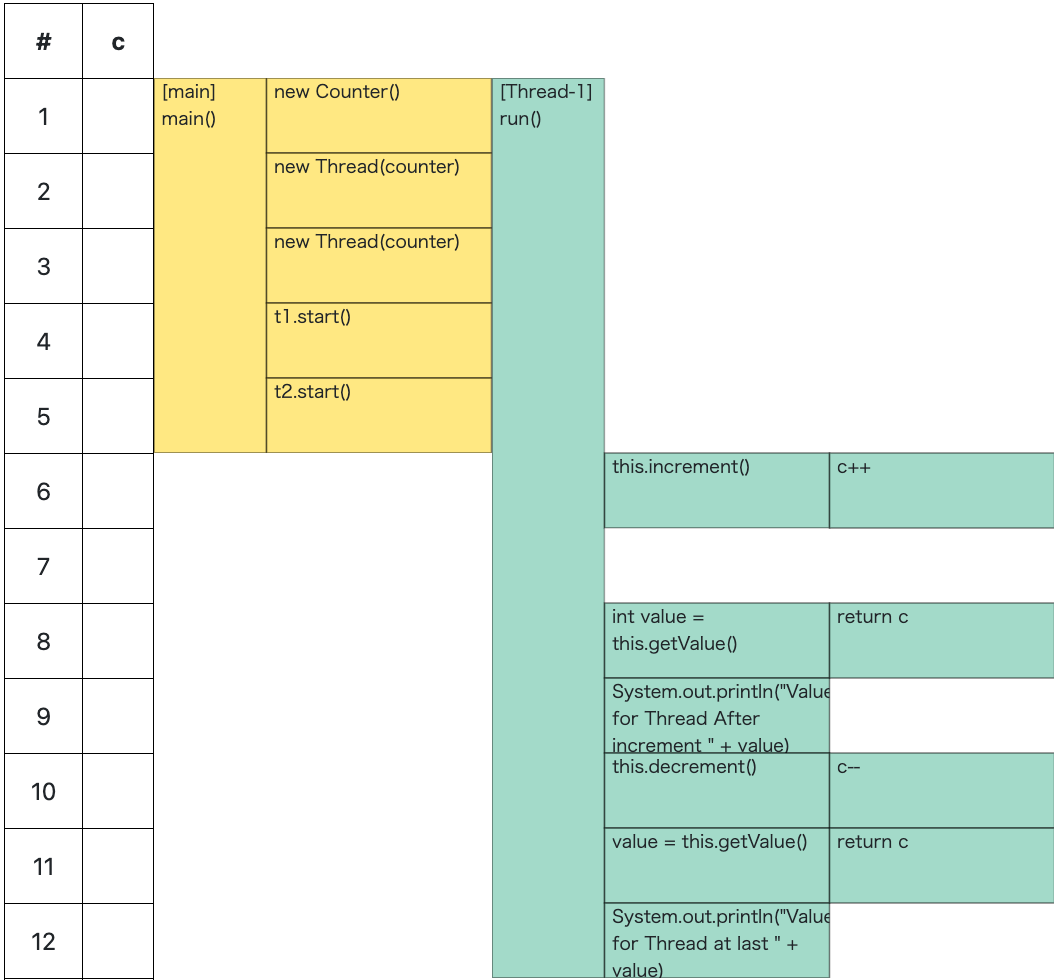}
        \caption{Part of the enhanced trace table}\label{拡張トレース表}
    \end{center}
\end{figure}

In this study, we use an enhanced trace table that combines traces and variable value input fields to facilitate visual understanding of the execution order between threads.
Listings \ref{ソースコード} and \ref{トレース} show a program that increments a counter using multiple threads and its trace, while Figure \ref{拡張トレース表} presents an enhanced trace table using that traces.
This program creates two threads, each of which increments the counter, outputs the counter value, then decrements the counter, and outputs the counter value again.
The correct behavior, as shown in Listing \ref{正しい出力}, is for the increments and decrements to alternate.
However, in practice, as shown in Listing \ref{実際の出力}, two increments are performed before the first output is made.
\begin{figure}[h]
    \begin{CJK}{UTF8}{ipxg}
    \begin{lstlisting}[caption=Correct output of counter program,label=正しい出力,escapechar=!]
      Value for Thread After increment 1
      Value for Thread at last 0
      Value for Thread After increment 1
      Value for Thread at last 0
    \end{lstlisting}
    \end{CJK}
\end{figure}
\begin{figure}[h]
    \begin{CJK}{UTF8}{ipxg}
    \begin{lstlisting}[caption=Output when a race condition occurs in the counter program,label=実際の出力,escapechar=!]
      Value for Thread After increment 2
      Value for Thread at last 1
      Value for Thread After increment 1
      Value for Thread at last 0
    \end{lstlisting}
    \end{CJK}
\end{figure}

The three-colored graph displayed on the right side of Figure \ref{拡張トレース表} represents the trace that records events occurring within the three threads.
This graph corresponds to the procedure numbers in the traditional trace table, and the values of the specified variables are entered in the input fields at the same height.
Nodes with the same thread information are represented by different colors: the \texttt{main} thread is represented by yellow nodes, and \texttt{thread-1} by green nodes.
The color-coded graph represents the tree structure of the event logs, with the largest yellow root node representing the \texttt{main()} method of the main thread.
To its right, the methods called within the \texttt{main()} method are displayed in the order of their recording from top to bottom.
Similarly, if method calls are nested deeper, new nodes are drawn to the right of the nodes to which the events belong.
Root nodes other than the \texttt{main} thread represent the \texttt{run()} methods.
In addition, empty spaces are drawn in the graph to indicate the periods when no other threads are running while instructions are being executed in a certain thread, thereby representing the sequence of event logs between threads.
\begin{figure}[h]
    \begin{CJK}{UTF8}{ipxg}
    \begin{lstlisting}[caption=Source code of counter program,label=ソースコード,escapechar=!]
      public class Main {
        public static void main(String args[]) {
            Counter counter = new Counter();
            Thread t1 = new Thread(counter);
            Thread t2 = new Thread(counter);
            t1.start();
            t2.start();
        }
      }
      class Counter implements Runnable {
        private int c = 0;
        public void increment() { c++; }
        public void decrement() { c--; }
        public int getValue() { return c; }
        @Override
        public void run() {
            this.increment();
            int value = this.getValue();
            System.out.println("Value for Thread After increment " + value);
            this.decrement();
            value = this.getValue();
            System.out.println("Value for Thread at last " + value);
        }
      }
    \end{lstlisting}
    \end{CJK}
\end{figure}
\begin{figure}[h]
    \begin{CJK}{UTF8}{ipxg}
    \begin{lstlisting}[caption=Trace of counter program,label=トレース,escapechar=!]
      [main] Counter counter = new Counter()
      [main] Thread t1 = new Thread(counter)
      [main] Thread t2 = new Thread(counter)
      [main] t1.start()
      [main] t2.start()
      [thread-1] this.increment()
      [thread-1] c++
      [thread-2] this.increment()
      [thread-2] c++
      [thread-1] int value = this.getValue()
      [thread-1] return c
      [thread-1] System.out.println("Value for Thread After increment " + value)
      [thread-1] this.decrement()
      [thread-1] c--
      [thread-1] value = this.getValue()
      [thread-1] return c
      [thread-1] System.out.println("Value for Thread at last " + value)
      [thread-2] int value = this.getValue()
      [thread-2] return c
      [thread-2] System.out.println("Value for Thread After increment " + value)
      [thread-2] this.decrement()
      [thread-2] c--
      [thread-2] value = this.getValue()
      [thread-2] return c
      [thread-2] System.out.println("Value for Thread at last " + value)
    \end{lstlisting}
    \end{CJK}
\end{figure}

\section{Implementation}
This study proposes a tool for learning multi-threaded behavior. 
With this tool, learners solve trace selection questions and variable value fill-in-the-blank questions to study the execution order of instructions in each thread of a multi-threaded program and the transitions of variable values accessed by multiple threads.
First, learners enter their names and select their learning tasks on the home screen.
Then, they proceed with two tasks: a selection question and a fill-in-the-blanks question.
After answering each question, the tool moves to an answer-checking screen, and if the answer is incorrect, the learner is prompted to answer again.

\subsection{Selection Question}
In the trace selection question, the learner selects the trace that corresponds to the given source code and its output.
Four choices are presented, with only one correct answer. Each choice differs in the order in which events are recorded.
The learner selects their answer from a drop-down list and clicks the confirm button to check their answer on the replay screen.
\begin{figure}[h]
    \begin{center}
        \includegraphics[width=7cm]{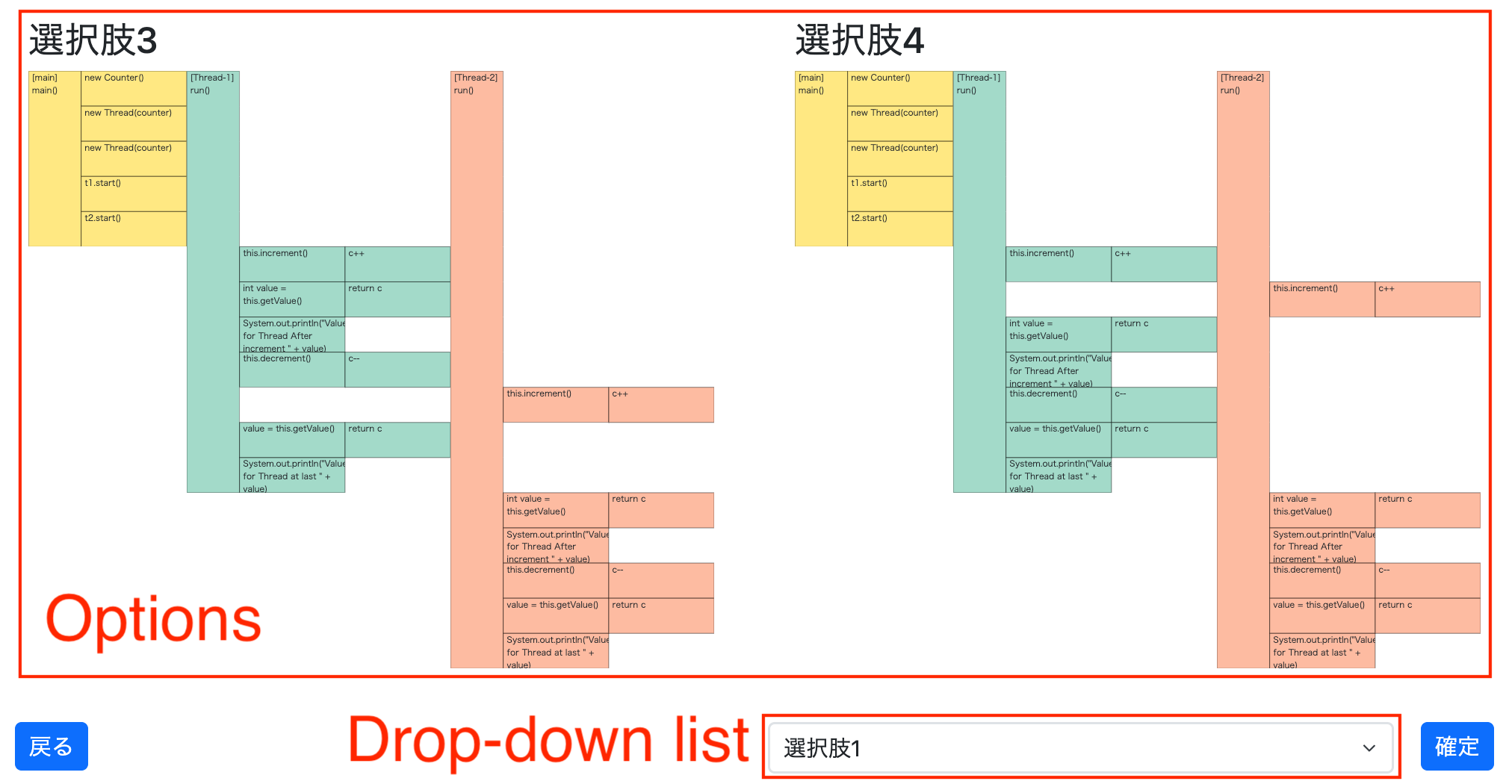}
        \caption{Part of selection question screen}\label{トレース選択問題画面画像}
    \end{center}
\end{figure}

\subsection{Answer Checking (Selection Question)}
On the answer checking screen for the selection question, the steps corresponding to the selected trace from the trace selection question screen are replayed, and the answer is checked.
In addition to indicating whether the answer to the trace selection question is correct or incorrect, the selected trace, the source code, and replay buttons are displayed. 
When the page is accessed, the first line of the \texttt{main()} method is highlighted.
By clicking the forward or backward buttons, the highlight moves through the source code in the same order as the trace. 
Additionally, the part of the trace corresponding to the currently highlighted source code is emphasized in red.
This allows the learner to confirm which part of the trace corresponds to the source code. 
If the selected answer is correct, clicking the "Next" button will proceed to the variable value fill-in-the-blank question.
If the answer is incorrect, the learner must attempt the trace selection question again.
\begin{figure}[h]
    \begin{center}
        \includegraphics[width=7cm]{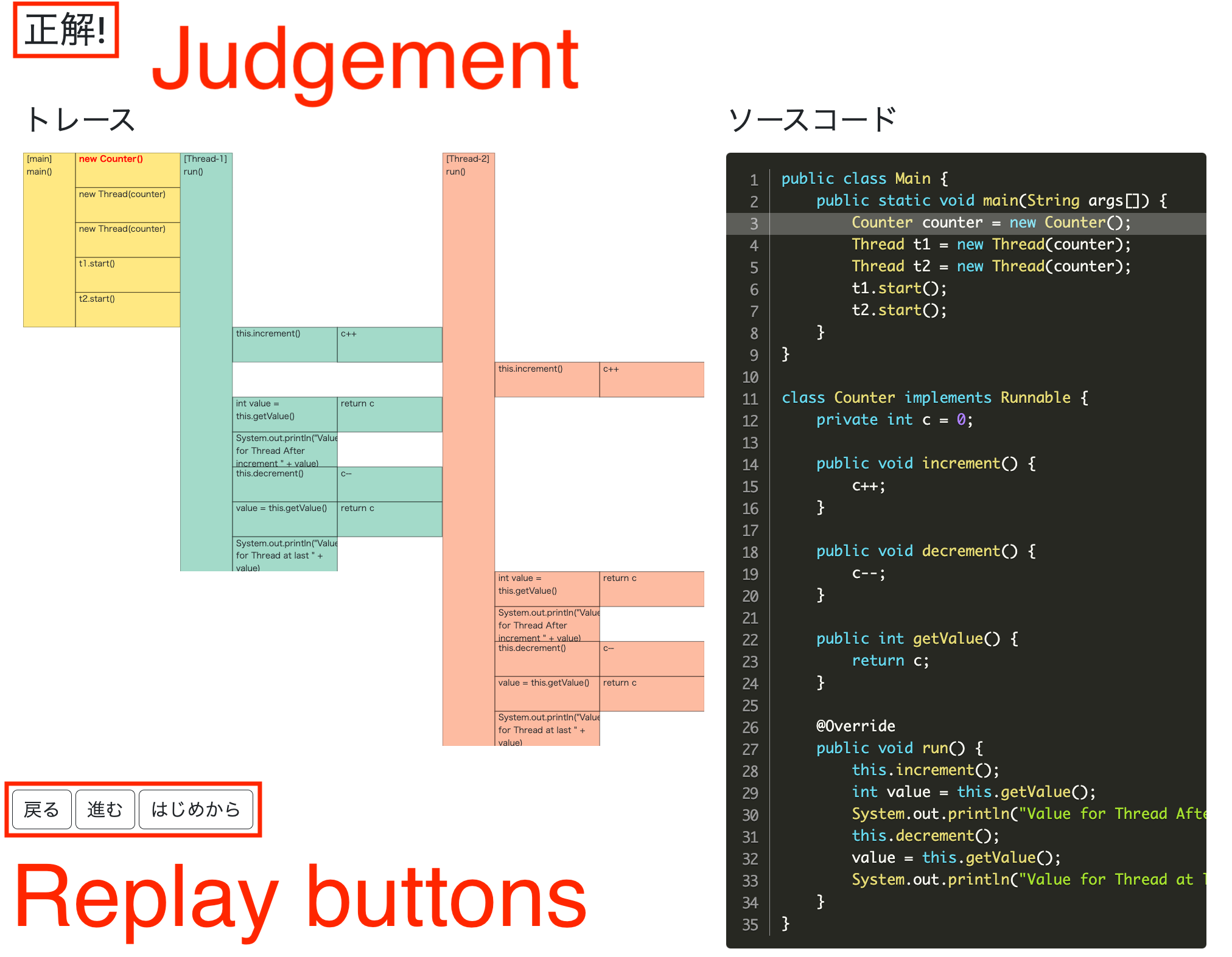}
        \caption{Answer checking screen (selection question)}\label{選択問題答え合わせ}
    \end{center}
\end{figure}

\subsection{Fill-in-the-blank}
The variable value fill-in-the-blank question involves entering the transitions of specified variable values.
The variables specified here are those involved in race conditions, accessed by multiple threads, and affect the output.
The enhanced trace table is displayed on the screen, and learners input the values immediately after the events recorded on the same row.
Once the learner has completed the input, they click the "Next" button to proceed to the results screen.
\begin{figure}[h]
    \begin{center}
        \includegraphics[width=7cm]{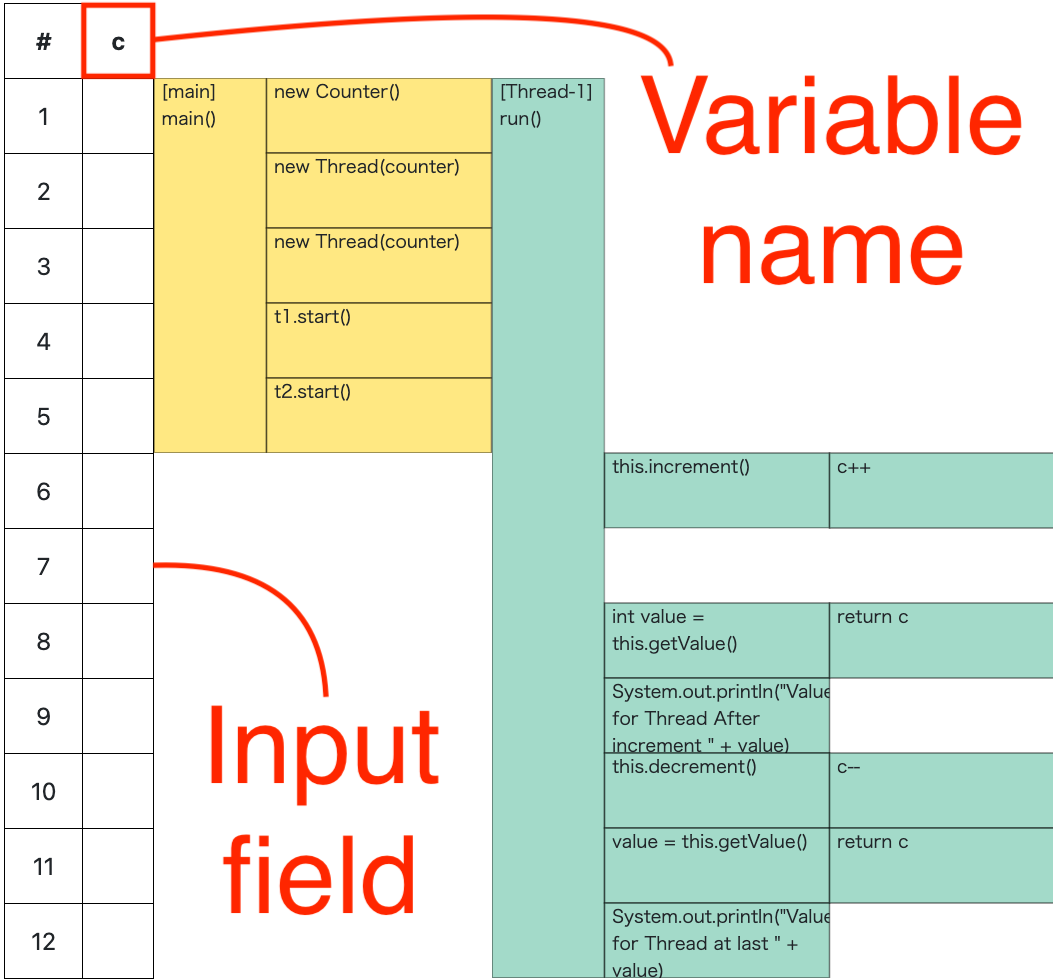}
        \caption{Part of fill-in-the-blank screen}\label{変数値穴埋め問題画像}
    \end{center}
\end{figure}

\subsection{Answer Checking (Fill-in-the-blank)}
On the answer checking screen for the fill-in-the-blank question, the correctness of the variable value fill-in-the-blank question is determined, and each input field is checked.
If the answers entered by the learner upon accessing the page are incorrect or left blank, they are highlighted in red.
Additionally, as a hint, the events that update the specified variable values are emphasized in red.
If all the input values are correct, the learning session is completed.
If any red input fields remain, the learner must re-enter the answers and click the "Confirm" button to check the answers again.
\begin{figure}[h]
    \begin{center}
        \includegraphics[width=7cm]{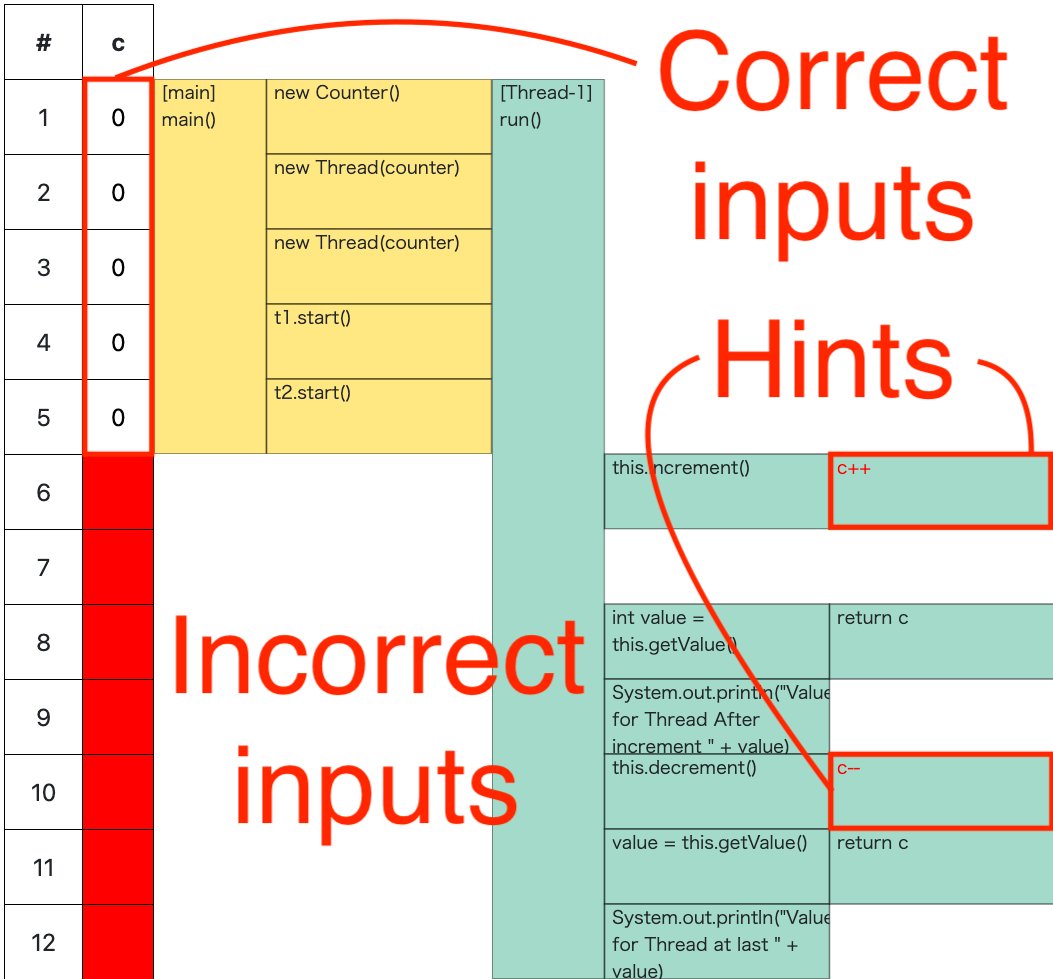}
        \caption{Answer checking screen (fill-in-the-blank)}\label{結果画面画像}
    \end{center}
\end{figure}

\section{Evaluation}
In this study, experiments were conducted to verify the following research questions (RQs).
\begin{enumerate}
    \item Is there a difference in test results before and after using the tool?
    \item Are any differences in the learning outcomes between text format traces and enhanced trace tables?
    \item What are the differences in learning between textual traces and enhanced trace tables?
\end{enumerate}

\subsection{Experiment}
The experiment was conducted with eight undergraduate students and two graduate students who had learned Java.
The participants were divided into two groups: the experimental group, which used a tool with an enhanced trace table, and the control group, which used a tool with a text-based trace.
The procedure of the experiment was as follows:
\begin{enumerate}
    \item Explain the entire experiment and the test.
    \item Conduct the pre-test.
    \item Explain the learning tool.
    \item Conduct the study twice.
    \item Conduct the post-test.
\end{enumerate}

\subsection{Pre-test and Post-test}
\begin{figure}[h]
    \begin{center}
        \includegraphics[width=7cm]{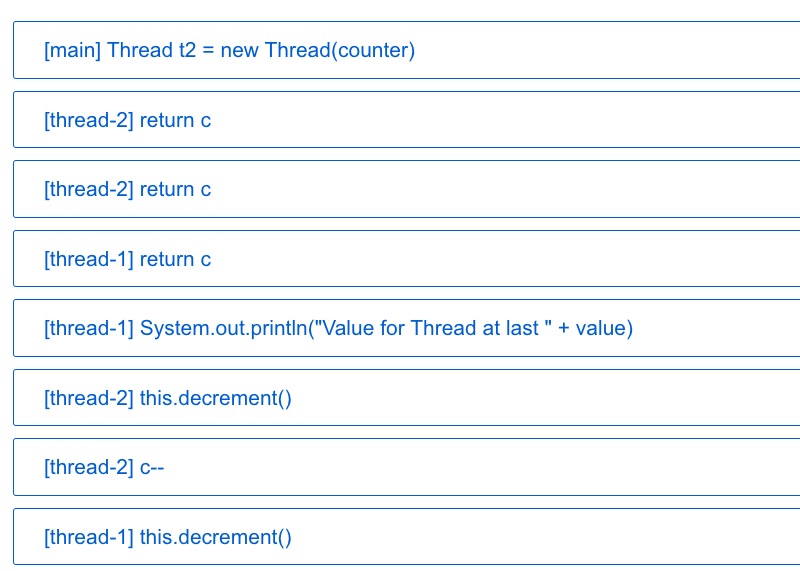}
        \caption{Pre-test and post-test}\label{事前事後テスト}
    \end{center}
\end{figure}

In the pre-test and post-test, a text-based trace ordering problem was conducted to measure the participants' understanding of multi-threaded behavior, as shown in Figure \ref{事前事後テスト}.
First, a printout containing the program's source code and the output produced when it is executed is distributed.
All of the programs are multi-threaded programs that perform standard output and have the potential to cause race conditions.
In addition, thread switching can occur at any point in the program.
Participants are asked to reproduce the trace recorded when the given output was generated by rearranging the provided choices.
The test continues until the participants press the "Grade" button.

The test results are evaluated by counting the number of choices placed in incorrect positions.
First, it is checked whether the events are arranged in an impossible order (execution order violation).
By separating the traces by thread and investigating them sequentially from the earliest trace, the test checks for traces that are placed even though the prerequisite events have not been recorded.
Next, the order in which the values of variables involved in the output are retrieved and updated is checked (retrieval-update order violation).
Similarly, by investigating the traces in order from the earliest one, the test compares them to the predefined retrieval-update order to check for traces that are placed without the prerequisite events being recorded.
If the traces are arranged in violation of the execution order, the number of execution order violations is counted as the number of incorrect choices.
If there are no execution order violations, but there are violations of the retrieval-update order, the number of retrieval-update order violations is counted as the number of incorrect choices.
Then, the number of errors is divided by the total number of choices to calculate the accuracy rate of the test.

\subsection{Control Group}
The control group learns using a learning tool that employs a text-based trace (Listing \ref{トレース}) instead of the enhanced trace table.
While there are no differences in the content of the tasks or the functionality of the learning tool compared to the tool used by the experimental group, the text-based trace is displayed in place of the enhanced trace table.

\section{Results}
\begin{figure}[h]
    \begin{center}
        \includegraphics[width=7cm]{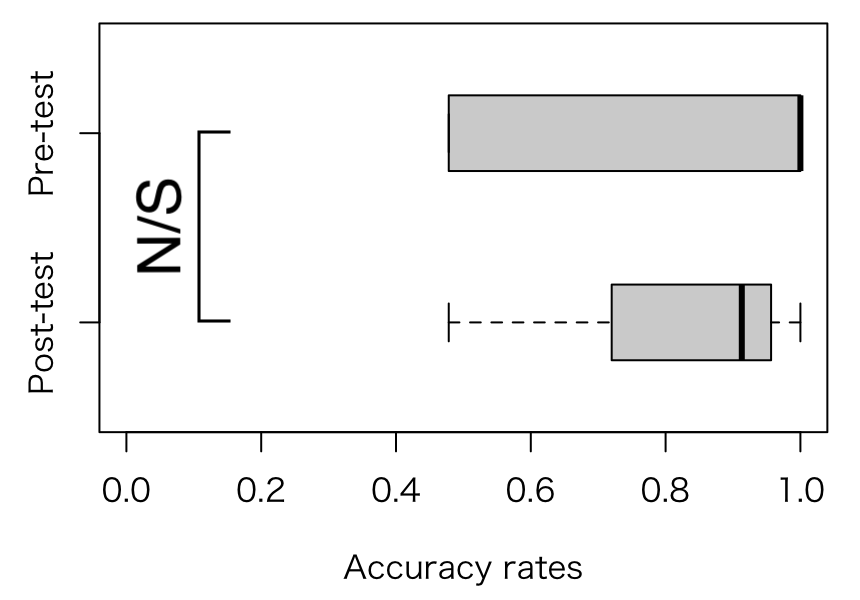}
        \caption{Comparison of pre-test and post-test of experimental groups}\label{実験群事前事後比較}
    \end{center}
\end{figure}

\begin{figure}[h]
    \begin{center}
        \includegraphics[width=7cm]{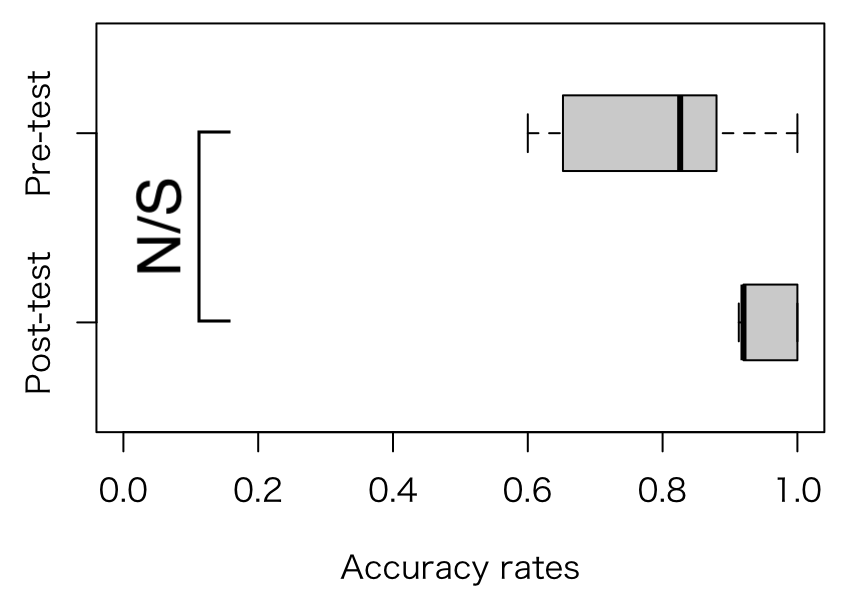}
        \caption{Comparison of pre-test and post-test of control groups}\label{統制群事前事後比較}
    \end{center}
\end{figure}

Figure \ref{実験群事前事後比較} compares the accuracy rates of the pre-test and post-test for the experimental group.
To determine whether there was a difference between the results of the pre-test and post-test, a paired t-test was conducted.
The results showed no significant difference ($p > 0.05$, $\alpha = 0.05$, $p = 0.7169$).
Figure \ref{統制群事前事後比較} compares the accuracy rates of the pre-test and post-test for the control group.
A paired t-test was conducted to assess differences between the pre-test and post-test results, and no significant difference was found ($p > 0.05$, $\alpha = 0.05$, $p = 0.1044$).
Therefore, no difference was found in the test results before and after learning with the tool.

\begin{figure}[h]
    \begin{center}
        \includegraphics[width=7cm]{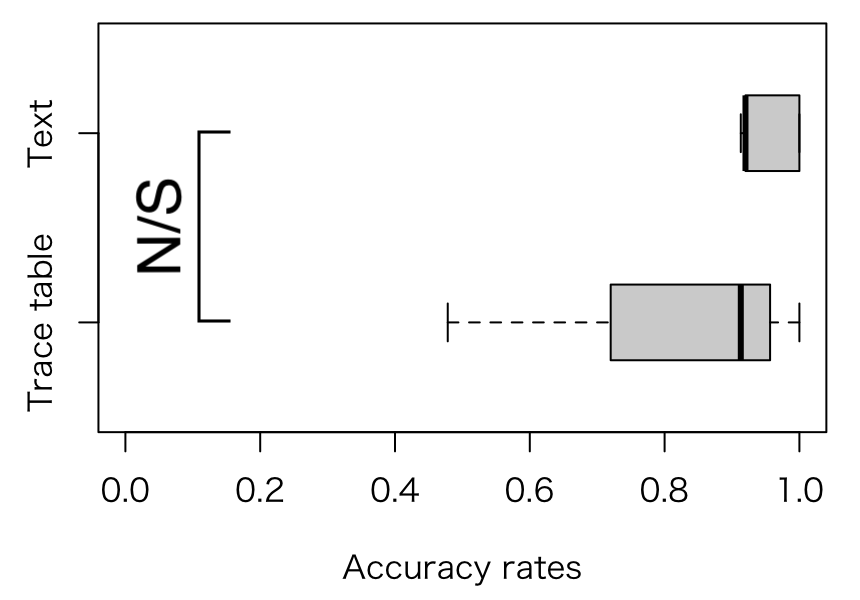}
        \caption{Comparison of post-test results between experimental and control groups}\label{テスト結果比較}
    \end{center}
\end{figure}

Figure \ref{テスト結果比較} compares the accuracy rates of the post-tests between the experimental group and the control group.
To determine whether there was a significant difference between the results of the experimental and control groups, the Mann-Whitney U test was conducted.
The result showed no significant difference ($p > 0.05$, $\alpha = 0.05$, $p = 0.3571$). 
Therefore, no difference in learning effectiveness was found between the text-based trace and the enhanced trace table.

\begin{figure}[h]
    \begin{center}
        \includegraphics[width=7cm]{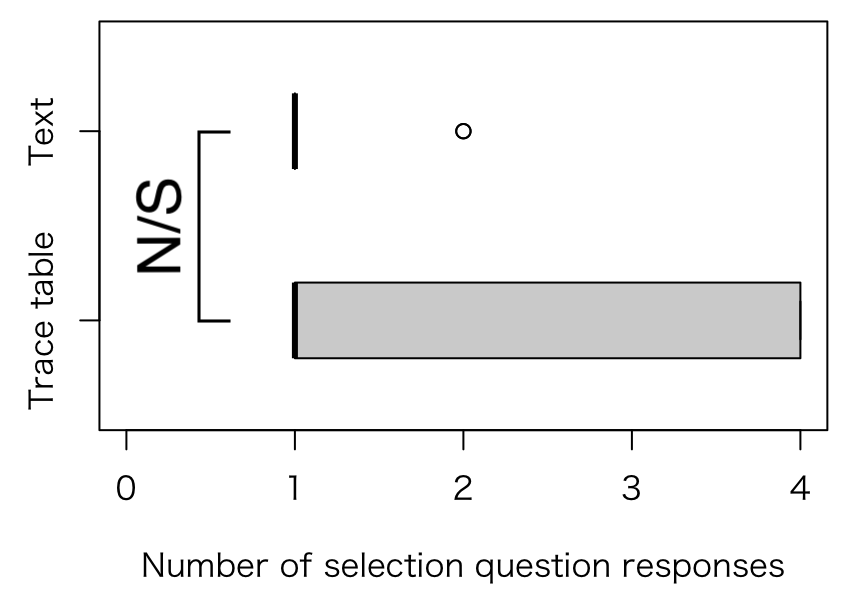}
        \caption{Comparison of the number of selection question responses between the experimental group and the control group}\label{選択問題回答回数}
    \end{center}
\end{figure}

\begin{figure}[h]
    \begin{center}
        \includegraphics[width=7cm]{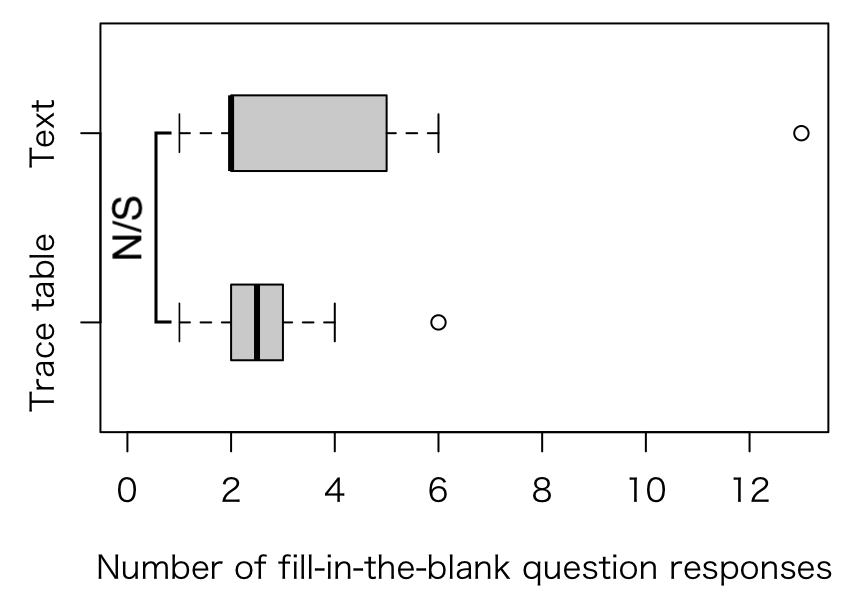}
        \caption{Comparison of the number of fill-in-the-blank question responses between the experimental group and the control group}\label{穴埋め問題回答回数}
    \end{center}
\end{figure}

Figure \ref{選択問題回答回数} compares of the number of selection question responses between the experimental group and the control group.
A Mann-Whitney U test was conducted to assess differences between the groups, and no significant difference was found ($p > 0.05$, $\alpha = 0.05$, $p = 0.3731$).
Figure \ref{穴埋め問題回答回数} compares the number of attempts for the fill-in-the-blank question between the two groups.
The Mann-Whitney U test showed no significant difference between the groups ($p > 0.05$, $\alpha = 0.05$, $p = 0.9002$).

\begin{figure}[h]
    \begin{center}
        \includegraphics[width=7cm]{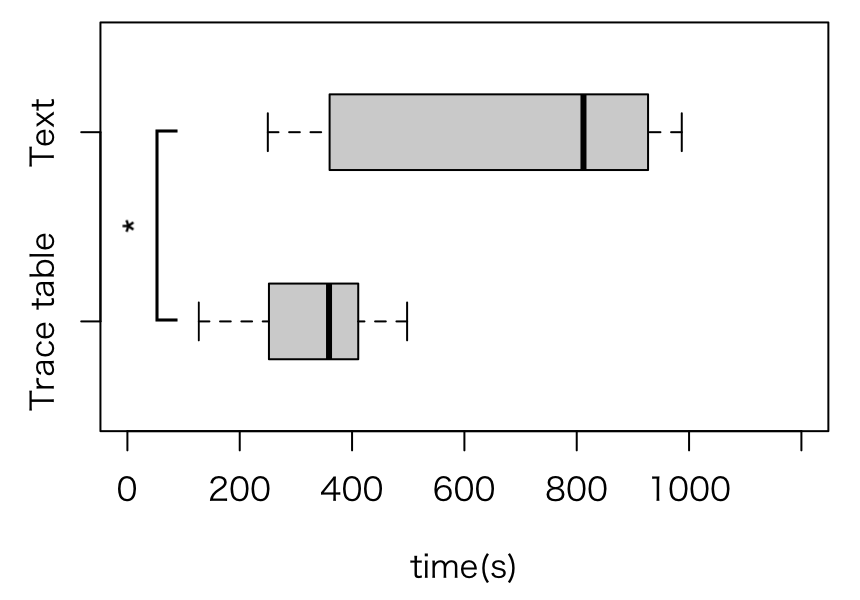}
        \caption{Comparison of learning time between experimental and control groups}\label{学習時間比較}
    \end{center}
\end{figure}

Figure \ref{学習時間比較} compares the learning time between the experimental and control groups.
An independent t-test was conducted to assess differences in learning times, and a significant difference was found ($p < 0.05$, $\alpha = 0.05$, $p = 0.002122$).
The average time taken for the experimental group was 341.3 seconds, whereas the control group took an average of 690.3 seconds.
This suggests that there is a difference in learning time between the text-based trace and the enhanced trace table.

Based on the above, the answers to the research questions (RQs) are as follows:
Regarding RQ1, a comparison of the accuracy rates between the pre-test and post-test for the experimental group showed no significant difference, suggesting no difference in test results before and after using the tool for learning.
Regarding RQ2, a comparison of the post-test accuracy rates between the experimental and control groups showed no significant difference, suggesting no difference in learning effectiveness between the text-based trace and the enhanced trace table.
Regarding RQ3, the comparison of the number of question attempts and learning times between the experimental and control groups showed no significant difference in the number of attempts, but a significant difference in learning time.
This suggests that there is a difference in learning time between the text-based trace and the enhanced trace table.

\section{Related Work}
Various debugging education tools have been proposed previously. For example, Miljanovic and Bradbury \cite{miljanovic_robobug_2017} proposed RoboBug, a debugging learning tool that can be customized according to programming language and game level. RoboBug is a puzzle-based game designed to teach computer science students how to debug for the first time. As learners advance through levels, various tools are unlocked, and learners must use these unlocked tools before employing the bug-capturing tool to clear each level. RoboBug was found to help students achieve success in learning debugging. However, it was observed that some participants could not complete the game due to the lack of hints.

Lee et al. \cite{lee_principles_2014} proposed Gidget, a debugging learning game. In Gidget, players help a character by diagnosing and fixing defective code to achieve mission goals at each level. Debugging in the game is performed using features similar to actual breakpoints. The game covers specific programming concepts across seven units and 34 levels using an imperative language like Python. Gidget has been reported to have a positive impact on participants' programming skills. However, it does not support the learning of multi-threaded programming concepts.

Various methods have also been proposed for trace visualization.
Malnati et al. \cite{malnati_jthreadspy_2007} proposed JThreadSpy, a trace recording and visualization tool for each thread, aimed at improving students’ understanding of multithreading. JThreadSpy records the method flow using dynamic instrumentation and displays the execution flow for each thread using diagrams that extend UML. It shows method calls, object associations, and simultaneous access to critical sections. Malnati et al. demonstrated deadlock detection by running examples introduced in Java tutorials with JThreadSpy, and they confirmed the wait time for lock acquisition through the generated diagrams. The tool was also used in object-oriented programming courses to teach concurrent processing to engineering students, and it was regarded as a valuable tool for visualizing abstract concepts related to concurrent programming.
However, it assumes that the program adheres to the principles of encapsulation, and it only handles events related to method calls and returns.

Various trace-based debugging methods have also been proposed. Schulz and Bockisch \cite{redshell} developed RedShell, which extends Eclipse Java Development Tools (JDT) to enable reverse debugging. RedShell allows reverse debugging by storing the results of program operations in its own local variables, enabling reverse debugging through Eclipse JDT’s Inspect feature. However, with RedShell, breakpoints must be set before running the target program, and developers need to have a detailed understanding of the program’s internal structure in advance.

\section{Conclusion}
The experimental results suggest that the enhanced trace table may contribute to reducing the learning time for multi-threaded behavior. However, there was no significant difference in the test results before and after learning for the experimental group, nor was there a significant difference between the post-test results of the experimental and control groups. This may be due to the fact that many participants spent time contemplating the problems on the problem screen without transitioning to the answer-checking screen, thus failing to utilize the hint function provided. Future challenges include improving the proposed method and conducting evaluations in larger-scale experiments.

\section*{Acknowledgment}	
Part of this research was supported by JSPS KAKENHI Grant Number 24K15214.

\bibliographystyle{IEEEtran}
\bibliography{ref}


\end{document}